\documentclass[aps,prb,preprint,showpacs,amsmath,amssymb]{revtex4}

\usepackage{subfigure}
\usepackage{graphicx}

\begin{document}


\title{Time-Dependent Density Functional Theory with Ultrasoft
Pseudopotential: Real-Time Electron Propagation across Molecular
Junction}

\author{Xiaofeng Qian,$^{1}$ Ju Li,$^{2}$ Xi Lin,$^{1}$ Sidney
Yip$^{1}$}

\email{syip@mit.edu}

\affiliation{$^1$Department of Nuclear Science and Engineering and
Department of Materials Science and Engineering, Massachusetts
Institute of Technology, Cambridge, Massachusetts 02139, USA}

\affiliation{$^2$Department of Materials Science and Engineering, Ohio
State University, Columbus, Ohio 43210, USA}

\date{\today}

\begin{abstract}

A practical computational scheme based on time-dependent density functional
theory (TDDFT) and ultrasoft pseudopotential (USPP) is developed to
study electron dynamics in real time. A modified Crank-Nicolson
time-stepping algorithm is adopted, under planewave basis. The scheme
is validated by calculating the optical absorption spectra for sodium
dimer and benzene molecule. As an application of this USPP-TDDFT
formalism, we compute the time evolution of a test electron packet at
the Fermi energy of the left metallic lead crossing a
benzene-(1,4)-dithiolate junction. A transmission probability of
5-7\%, corresponding to a conductance of 4.0-5.6 $\mu$S, is
obtained. These results are consistent with complex band structure
estimates, and Green's function calculation results at small bias
voltages.

\end{abstract}


\maketitle

\section{Introduction}

The development of molecular scale electronic devices has attracted a
great deal of interest in the past decade, although major experimental
and theoretical challenges still
exist. \cite{ReedZMBT97,SmitNULvv02,JoachimGA00,Nitzan01,HeathR03} To
date precise experimental control of molecular conformation is
lacking, resulting in large uncertainties in the measured
conductance. On the theory side, while the Green's function (GF)
method has achieved many successes in describing electron transport at
the meso \cite{Datta95,Imry02} and molecular
\cite{DerosaS01,DamleGD01,TaylorGW01,XueR03,KeBY04} scales, issues
such as dynamical electron correlation and large electron-phonon
coupling effects \cite{GalperinRN05,GalperinNRS05} are far from fully
resolved. It is therefore desirable to exploit alternative approaches
for comparison with the mainstream GF calculations. In this paper, we
describe a first step towards this goal by computing how an electron
propagates through a molecular junction in real time, based on the
time-dependent density functional theory \cite{RungeG84} (TDDFT).

Density functional theory (DFT) \cite{HohenbergK64} with the Kohn-Sham
reference kinetic energy functional of a fictitious non-interacting
electron system \cite{KohnS65} is a leading method for treating many
electrons in solids and molecules.  \cite{ParrY89}. While initially
formulated to describe only the electronic ground state
\cite{HohenbergK64,KohnS65}, it has been rigorously extended by Runge
and Gross \cite{RungeG84} to treat time-dependent, driven systems
(excited states). TDDFT is therefore a natural theoretical platform
for studying electron conduction at the nanoscale. There are two
flavors in which TDDFT is implemented. One is direct numerical
integration
\cite{YabanaB96,YabanaB99,BertschIRY00,MarquesCBR03,TsolakidisSM02,CastroMR04}
of the time-dependent Kohn-Sham (TDKS) equations. The other is a {\em
Gedanken experiment} of the former with an added assumption of
infinitesimal time-dependent perturbation, so a linear response
function may be first derived in closed form
\cite{BauernschmittA96,CasidaJCS98,ChelikowskyKV03}, which is then
evaluated numerically. These two implementations should give exactly
the same result when the external perturbation field is
infinitesimal. The latter implementation can be computationally more
efficient once the linear-response function has been analytically
derived, while the former can treat non-infinitesimal perturbations
and arbitrary initial states.

A key step of the TDDFT dynamics is updating of the Kohn-Sham
effective potential by the present {\em excited-state} charge density
$\rho({\bf x},t)$, $\hat{V}_{\rm KS}(t)=\hat{V}_{\rm KS}[\rho({\bf
x},t),...]$.  This is what sets TDDFT apart from the ground-state DFT
estimate of excitation energies, even when TDDFT is applied in its
crudest, so-called adiabatic approximation, \cite{BauernschmittA96}
whereby the same exchange-correlation density functional form as the
ground-state DFT calculation is used (for example, the so-called TDLDA
approximation uses exactly the same Ceperley-Alder-Perdew-Zunger
functional \cite{CeperleyA80,PerdewZ81} as the ground-state LDA
calculation.)  This difference in excitation energies comes about
because in a ground-state DFT calculation, a virtual orbital such as
LUMO (lowest unoccupied molecular orbital) experiences an effective
potential due to $N$ electrons occupying the lowest $N$ orbitals;
whereas in a TDDFT calculation, if one electron is excited to a
LUMO-like orbital, it sees $N-1$ electrons occupying the lowest $N-1$
orbitals, plus its own charge density. Also, the excitation energy is
defined by the collective reaction of this coupled dynamical system to
time-dependent perturbation (pole in the response function)
\cite{LiY97}, rather than simple algebraic differences between present
virtual and occupied orbital energies. For rather involved reasons
beyond what is discussed here, TDDFT under the adiabatic approximation
gives significantly improved excitation spectra
\cite{BauernschmittA96,CasidaJCS98}, although there are still much to
be desired. Further systematic improvements to TDDFT such as current
density functional \cite{VignaleK96} and self-interaction correction
\cite{TongC98} have already made great strides.

Presently, most electronic conductance calculations based on the
Landauer transmission formalism \cite{Landauer57,Landauer70} have
assumed a static molecular geometry. In the Landauer picture,
dissipation of the conducting electron energy is assumed to take place
in the metallic leads (electron reservoirs), not in the narrow
molecular junction (channel) itself. \cite{ImryL99} Inelastic
scattering, however, does occur in the molecular junctions themselves,
the effects appearing as peaks or dips in the measured inelastic
electron tunneling spectra (IETS) \cite{GalperinRN04} at molecular
vibrational eigen-frequencies. Since heating is always an important
concern for high-density electronics, and because molecular junctions
tend to be mechanically more fragile compared to larger,
semiconductor-based devices, the issue of electron-phonon coupling
warrants detailed calculations \cite{GalperinRN04,FrederiksenBLJ04}
(here we use the word phonon to denote general vibrations when there
is no translational symmetry).  In the case of long $\pi$-conjugated
polymer chain junctions, strong electron-phonon coupling may even lead
to new elementary excitations and spin or charge carriers, called
soliton/polaron
\cite{HeegerKSS88,GalperinRN05,GalperinNRS05,LinLSY05,LinLS05}, where
the electronic excitation is so entangled with phonon excitation that
separation is no longer possible.

In view of the above background, there is a need for efficient TDDFT
implementations that can treat complex electron-electron and
electron-phonon interactions in the time domain.  Linear-response type
analytic derivations can become very cumbersome, and for some problems
\cite{CalvayracRSU00} may be entirely infeasible. A direct
time-stepping method
\cite{YabanaB96,YabanaB99,BertschIRY00,MarquesCBR03,CastroMR04,TsolakidisSM02}
analogous to molecular dynamics for electrons as well as ions may be
more flexible and intuitive in treating some of these highly complex
and coupled problems, {\em if} the computational costs can be
managed. Such a direct time-stepping code also can be used to
double-check the correctness of analytic approaches such as the
non-equilibrium Green's function (NEGF) method and electron-phonon
scattering calculations \cite{GalperinRN04,FrederiksenBLJ04}, most of
which explicitly or implicitly use the same set of TDDFT
approximations (most often an adiabatic approximation such as TDLDA).

Two issues are of utmost importance when it comes to computational
cost: choice of basis and pseudopotential. For ground-state DFT
calculations that involve a significant number of metal atoms
(e.g. surface catalysis), the method that tends to achieve the best
cost-performance compromise is the ultrasoft pseudopotential (USPP)
\cite{Vanderbilt90,LaasonenCLV91,LaasonenPCLV93} with planewave basis,
and an independent and theoretically more rigorous formulation, the
projector augmented-wave (PAW) \cite{Blochl94} method. Compared to the
more traditional norm-conserving pseudopotential approaches, USPP/PAW
achieve dramatic cost savings for first-row $p$- and $d$-elements,
with minimal loss of accuracy. USPP/PAW are the workhorses in popular
codes such as VASP \cite{KresseF96} and DACAPO
\cite{DACAPO,HammerHN99,BahnJ02}. We note that similar to surface
catalysis problems, metal-molecule interaction at contact is the key
for electron conduction across molecular junctions. Therefore it seems
reasonable to explore how TDDFT, specifically TDKS under the adiabatic
approximation, performs in the USPP/PAW framework, which may achieve
similar cost-performance benefits. This is the main distinction
between our approach and the software package Octopus
\cite{MarquesCBR03,CastroMR04}, a ground-breaking TDDFT program with
direct time stepping, but which uses norm-conserving Troullier-Martins
(TM) pseudopotential \cite{TroullierM91}, and real-space grids. We
will address the theoretical formulation of TD-USPP (TD-PAW) in
sec. II, and the numerical implementation of TD-USPP in the direct
time-stepping flavor in sec. III.

To validate that the direct time-integration USPP-TDDFT algorithm
indeed works, we calculate the optical absorption spectra of sodium
dimer and benzene molecule in sec. IV and compare them with
experimental results and other TDLDA calculations. As an application,
we perform a computer experiment in sec. V which is a verbatim
implementation of the original Landauer picture
\cite{Landauer70,ImryL99}. An electron wave pack comes from the left
metallic lead (1D Au chain) with an energy that is exactly the Fermi
energy of the metal (the Fermi electron), and undergoes scattering by
the molecular junction (benzene-(1,4)-dithiolate, or BDT).  The
probability of electron transmission is carefully analyzed in density
vs. ${\bf x},t$ plots. The point of this exercise is to check the
stability and accuracy of the time integrator, rather than to obtain
new results about the Au-BDT-Au junction conductance. We check the
transmission probability thus obtained with simple estimate from
complex band structure calculations \cite{TomfohrS02a,TomfohrS02b},
and Green's function calculations at small bias voltages. Both seem to
be consistent with our calculations. Lastly, we give a brief summary
in sec. VI.

\section{TDDFT formalism with ultrasoft pseudopotential}

The key idea of USPP/PAW
\cite{Vanderbilt90,LaasonenCLV91,LaasonenPCLV93,Blochl94} is a mapping
of the true valence electron wavefunction $\tilde{\psi}({\bf x})$ to a
pseudowavefunction $\psi({\bf x})$: $\tilde{\psi}\leftrightarrow
\psi$, like in any pseudopotential scheme. However, by discarding the
requirement that $\psi({\bf x})$ must be norm-conserved ($\langle
\psi| \psi\rangle=1$) while matching $\tilde{\psi}({\bf x})$ outside
the pseudopotential cutoff, a greater smoothness of $\psi({\bf x})$ in
the core region can be achieved; and therefore less planewaves are
required to represent $\psi({\bf x})$. In order for the physics to
still work, one must define augmentation charges in the core region,
and solve a generalized eigenvalue problem
\begin{equation}
 \hat{H} |\psi_{n}\rangle = \varepsilon_{n} \hat{S} |\psi_{n}\rangle,
 \label{USPP-KS}
\end{equation}
instead of the traditional eigenvalue problem, where $\hat{S}$ is a
Hermitian and positive definite operator. $\hat{S}$ specifies the
fundamental measure of the linear Hilbert space of
pseudowavefunctions.  Physically meaningful inner product between two
pseudowavefunctions is always $\langle \psi| \hat{S} |
\psi^\prime\rangle$ instead of $\langle \psi|\psi^\prime\rangle$. For
instance, $\langle \psi_m | \psi_n\rangle\neq \delta_{mn}$ between the
eigenfunctions of (\ref{USPP-KS}) because it is actually not
physically meaningful, but $\langle \psi_m | \hat{S} | \psi_n\rangle
\equiv \langle \tilde{\psi}_m | \tilde{\psi}_n\rangle = \delta_{mn}$
is.  (Please note that $\tilde{\psi}$ is used to denote the true
wavefunction with nodal structure, and ${\psi}$ to denote
pseudowavefunction, which are opposite in some papers.)


$\hat{H}$ consists of the kinetic energy operator $\hat{T}$, ionic
local pseudopotential $\hat{V}_{\rm L}$, ionic nonlocal
pseudopotential $\hat{V}_{\rm NL}$, Hartree potential $\hat{V}_{\rm
H}$, and exchange-correlation potential $\hat{V}_{\rm XC}$,
\begin{equation}
\hat{H}
 = \hat{T} + \hat{V}_{\rm L} + \hat{V}_{\rm NL} + \hat{V}_{\rm
  H} + \hat{V}_{\rm {XC}}.
\end{equation}
The $\hat{S}$ operator is given by
\begin{equation}
\hat{S} = 1 + \sum_{i,j,I} q_{ij}^{I}
|\beta_{j}^{I}\rangle\langle\beta_{i}^{I}|,
\end{equation}
where $i\equiv(\tau lm)$ is the angular momentum channel number, and
$I$ labels the ions.  $\hat{S}$ contains contributions from all ions
in the supercell, just as the total pseudopotential operator
$\hat{V}_{\rm L}+\hat{V}_{\rm NL}$, which is the sum of
pseudopotential operators of all ions. In above, the projector
function $\beta_{i}^{I}({\bf x})\equiv \langle {\bf
x}|\beta_{i}^{I}\rangle$ of atom $I$'s channel $i$ is
\begin{equation}
\beta_{i}^{I}({\bf x})
=\beta_{i} ({\bf x}-{\bf X}_I),
\end{equation}
where ${\bf X}_I$ is the ion position, and $\beta_{i}({\bf x})$
vanishes outside the pseudopotential cutoff. These projector functions
appear in the nonlocal pseudopotential
\begin{equation}
\hat{V}_{\rm NL} = \sum_{i,j,I} D_{ji}^{I}
|\beta_{j}^{I}\rangle\langle\beta_{i}^{I}|,
\end{equation}
as well, where
\begin{equation}
D_{ji}^{I} = D_{ji}^{I(0)} + \int  d{\bf x} ({V}_{\rm L}({\bf x}) +
{V}_{\rm H}({\bf x}) + {V}_{\rm {XC}}({\bf x}))  Q_{ji}^{I}({\bf x}).
\end{equation}
The coefficients $D_{ji}^{I(0)}$ are the unscreened scattering
strengths, while the coefficients $D_{ji}^{I}$ need to be
self-consistently updated with the electron density
\begin{equation}
\rho({\bf x}) = \sum_{n} \left\{ \;|\psi_{n}|^2 + \sum_{i,j,I}
  Q_{ji}^I({\bf x}) \langle \psi_{n}|\beta_{j}^{I}\rangle
  \langle\beta_{i}^{I} |\psi_{n}\rangle \; \right\}
  f(\varepsilon_{n}),
\label{Q_charge}
\end{equation}
in which $f(\varepsilon_{n})$ is the Fermi-Dirac
distribution. $Q_{ji}^I({\bf x})$ is the charge augmentation function,
i.e., the difference between the true wavefunction charge
(interference) and the pseudocharge for selected channels,
\begin{equation}
  Q_{ji}^I({\bf x}) \;\equiv\; \tilde{\psi}_j^{I*}({\bf
 x})\tilde{\psi}_i^I({\bf x}) - \psi_j^{I*}({\bf x})\psi_i^I({\bf x}),
\end{equation}
which vanishes outside the cutoff. There is also
\begin{equation}
 q_{ij}^{I} \;\equiv\; \int d{\bf x} Q_{ji}^I({\bf x}).
\end{equation}
Terms in Eq. (\ref{Q_charge}) are evaluated using two different grids,
a sparse grid for the wavefunctions $\psi_{n}$ and a dense grid
for the augmentation functions $Q_{ji}^I({\bf x})$. Ultrasoft
pseudopotentials are thus fully specified by the functions $V_{\rm
L}({\bf x})$, $\beta_{i}^{I}({\bf x})$, $D_{ji}^{I(0)}$, and
$Q^{I}_{ji}({\bf x})$.  Forces on ions and internal stress on the
supercell can be derived analytically using linear response theory
\cite{LaasonenPCLV93,KresseF96}.

To extend the above ground-state USPP formalism to the time-dependent
case, we note that the $\hat{S}$ operator in (\ref{USPP-KS}) depends
on the ionic positions $\{{\bf X}_I\}$ only and {\em not} on the
electronic charge density. In the case that the ions are not moving,
the following dynamical equations are equivalent:
\begin{equation}
 \hat{H}(t) \psi_{n}(t)
= i\hbar \partial_t (\hat{S} \psi_{n}(t))
= \hat{S} (i\hbar\partial_t\psi_{n}(t)),
\end{equation}
whereby we have replaced the $\varepsilon_{n}$ in (\ref{USPP-KS}) by
the $i\hbar \partial_t$ operator, and $\hat{H}(t)$ is updated using
the time-dependent $\rho({\bf x},t)$. However when the ions are
moving,
\begin{equation}
  i\hbar \partial_t \hat{S} \;\neq\; \hat{S} (i\hbar\partial_t)
\end{equation}
with difference proportional to the ionic velocities. To resolve this
ambiguity, we note that $\hat{S}$ can be split as
\begin{equation}
 \hat{S} \;=\; (\hat{S}^{1/2}\hat{U})(\hat{U}^\dagger\hat{S}^{1/2}),
\end{equation}
where $\hat{U}$ is a unitary operator,
$\hat{U}\hat{U}^\dagger=\hat{I}$, and we can rewrite (\ref{USPP-KS})
as
\begin{equation}
 (\hat{U}^\dagger\hat{S}^{-1/2}) \hat{H}
(\hat{S}^{-1/2}\hat{U}) (\hat{U}^\dagger\hat{S}^{1/2})
\psi_{n} \;=\; \varepsilon_{n} (\hat{U}^\dagger\hat{S}^{1/2})
\psi_{n}.
 \label{USPP-Intermediate}
\end{equation}
Referring to the PAW formulation \cite{Blochl94}, we can select
$\hat{U}$ such that $\hat{U}^\dagger\hat{S}^{1/2}$ is the PAW
transformation operator
\begin{equation}
 \hat{U}^\dagger\hat{S}^{1/2} =
 \hat{T} \equiv 1+\sum_{i,I} (|\tilde{\psi}_i^I\rangle - |\psi_i^I\rangle)
 \langle \beta_i^I |: \;\;\;
 \tilde{\psi_{n}}=\hat{T}\psi_{n},
 \label{PAW_Transformation}
\end{equation}
that maps the pseudowavefunction to the true wavefunction. So we
can rewrite (\ref{USPP-Intermediate}) as,
\begin{equation}
  (\hat{U}^\dagger\hat{S}^{-1/2}) \hat{H} (\hat{S}^{-1/2}\hat{U})
\tilde{\psi_{n}} \;\equiv\; \hat{\tilde{H}} \tilde{\psi_{n}} \;=\;
\varepsilon_{n} \tilde{\psi_{n}},
\end{equation}
where $\hat{\tilde{H}}$ is then the true all-electron Hamiltonian
(with core-level electrons frozen). In the all-electron TDDFT
procedure, the above $\varepsilon_{n}$ is replaced by the $i\hbar
\partial_t$ operator. It is thus clear that a physically meaningful
TD-USPP equation in the case of moving ions should be
\begin{equation}
 (\hat{U}^\dagger\hat{S}^{-1/2}) \hat{H}
(\hat{S}^{-1/2}\hat{U}) (\hat{U}^\dagger\hat{S}^{1/2})
\psi_{n} \;=\; i\hbar \partial_t ((\hat{U}^\dagger\hat{S}^{1/2})
\psi_{n}),
\end{equation}
or
\begin{equation}
 (\hat{U}^\dagger\hat{S}^{-1/2}) \hat{H}
\psi_{n} \;=\; i\hbar \partial_t ((\hat{U}^\dagger\hat{S}^{1/2})
\psi_{n}).
\end{equation}
In the equivalent PAW notation, it is simply,
\begin{equation}
 (\hat{T}^\dagger)^{-1}\hat{H}\psi_{n} \;=\;
 i\hbar \partial_t (\hat{T} \psi_{n}).
\end{equation}
Or, in pseudized form amenable to numerical calculations,
\begin{equation}
 \hat{H}\psi_{n} =
 i\hbar \hat{T}^\dagger (\partial_t (\hat{T} \psi_{n})) = i\hbar
 (\hat{T}^\dagger\hat{T} (\partial_t \psi_{n}) +
 \hat{T}^\dagger(\partial_t \hat{T}) \psi_{n}).
\end{equation}
Differentiating (\ref{PAW_Transformation}), there is,
\begin{equation}
 \partial_t \hat{T} \;=\;
\sum_{i,I} \left(\frac{\partial (|\tilde{\psi}_i^I\rangle -
  |\psi_i^I\rangle)}
{\partial {\bf X}_I}
 \langle \beta_i^I | + (|\tilde{\psi}_i^I\rangle - |\psi_i^I\rangle)
 \frac{\partial \langle\beta_i^I|}{\partial {\bf X}_I}\right)\cdot
 \dot{\bf X}_I,
\end{equation}
and so we can define and calculate
\begin{equation}
 \hat{P} \;\equiv\; -i\hbar\hat{T}^\dagger(\partial_t \hat{T}) =
 \sum_{i,I} \hat{\bf P}^I \cdot \dot{\bf X}_I
 \label{Poperator}
\end{equation}
operator, similar to analytic force calculation
\cite{LaasonenPCLV93}. The TD-USPP / TD-PAW equation therefore can be
rearranged as,
\begin{equation}
 (\hat{H}+\hat{P})\psi_{n} \;=\;  i\hbar \hat{S} (\partial_t\psi_{n}),
 \label{TD-USPP-PAW}
\end{equation}
with $\hat{P}$ proportional to the ionic velocities. It is basically
the same as traditional TDDFT equation, but taking into account the
moving spatial ``gauge'' due to ion motion. As such it can be used to
model electron-phonon coupling \cite{FrederiksenBLJ04}, cluster
dynamics under strong laser field \cite{CalvayracRSU00}, etc., as long
as the pseudopotential cores are not overlapping, and the core-level
electrons are not excited.

At each timestep, one should update $\rho({\bf x},t)$ as
\begin{equation}
\rho({\bf x},t) = \sum_{n} \left\{ \;|\psi_{n}({\bf x},t)|^2 +
  \sum_{i,j,I} Q_{ji}^I({\bf x}) \langle
  \psi_{n}(t)|\beta_{j}^{I}\rangle \langle\beta_{i}^{I}
  |\psi_{n}(t)\rangle \; \right\} f_{n}.
 \label{TD-USPP-charge}
\end{equation}
Note that while $\psi_{n}({\bf x},t=0)$ may be an eigenstate if we
start from the ground-state wavefunctions, $\psi_{n}({\bf x},t>0)$
generally is no longer so with the external field turned on. $n$ is
therefore merely used as a label based on the initial state rather
than an eigenstate label at $t>0$. $f_{n}$ on the other hand always
maintains its initial value, $f_{n}(t)=f_{n}(0)$, for a particular
simulation run.

One may define projection operator $\hat{t}_I$ belonging to atom $I$:
\begin{equation}
 \hat{t}_I \;\equiv\; \sum_i (|\tilde{\psi}_i^I\rangle - |\psi_i^I\rangle)
 \langle \beta_i^I |.
\end{equation}
$\hat{t}_I$ spatially has finite support, and so is
\begin{equation}
 \frac{\partial \hat{t}_I}{\partial {\bf X}_I}
 = -\frac{\partial \hat{t}_I}{\partial {\bf x}}
 = -\frac{\partial(1+\hat{t}_I)}{\partial {\bf x}}
 = (1+\hat{t}_I) \nabla - \nabla (1+\hat{t}_I).
\end{equation}
Therefore $\hat{\bf P}^I$ in (\ref{Poperator}) is,
\begin{eqnarray}
 \hat{\bf P}^I \;\;=&& -i\hbar\hat{T}^\dagger
  \frac{\partial \hat{t}_I}{\partial {\bf X}_I} \nonumber\\
=&&  -i\hbar(1+\hat{t}_I^\dagger)
 \frac{\partial \hat{t}_I}{\partial {\bf X}_I} \nonumber\\
=&&  -i\hbar(1+\hat{t}_I^\dagger)((1+\hat{t}_I) \nabla - \nabla (1+\hat{t}_I))
\nonumber\\
=&& (1+\hat{t}_I^\dagger)(1+\hat{t}_I) {\bf p} -
    (1+\hat{t}_I^\dagger){\bf p}(1+\hat{t}_I),
\end{eqnarray}
where ${\bf p}$ is the electron momentum operator. Unfortunately
$\hat{\bf P}^I$ and therefore $\hat{P}$ are not Hermitian
operators. This means that the numerical algorithm for integrating
(\ref{TD-USPP-PAW}) may be different from the special case of immobile
ions:
\begin{equation}
 \hat{H}(t) \psi_{n} \;=\;  i\hbar \hat{S} (\partial_t\psi_{n}).
 \label{TD-USPP-Immobile}
\end{equation}
Even if the same time-stepping algorithm is used, the error estimates
would be different. In section III we discuss algorithms for
integrating (\ref{TD-USPP-Immobile}) only, and postpone detailed
discussion of integration algorithm and error estimates for coupled
ion-electron dynamics (\ref{TD-USPP-PAW}) under USPP to a later paper.

\section{Time-Stepping Algorithms for the Case of Immobile Ions}

In this section we focus on the important limiting case of
(\ref{TD-USPP-Immobile}), where the ions are immobile or can be
approximated as immobile. We may rewrite (\ref{TD-USPP-Immobile})
formally as
\begin{equation}
 \hat{S}^{-1/2}\hat{H}(t)\hat{S}^{-1/2} (\hat{S}^{1/2}\psi_{n}) \;=\;
 i\hbar \partial_t (\hat{S}^{1/2}\psi_{n}).
\end{equation}
And so the time evolution of (\ref{TD-USPP-Immobile}) can be formally
expressed as
\begin{equation}
 \psi_{n}(t) \;=\; \hat{S}^{-1/2}
 \hat{\cal T}\left[\exp\left(-\frac{i}{\hbar}
 \int_0^t dt^\prime
 \hat{S}^{-1/2}\hat{H}(t^\prime)\hat{S}^{-1/2}\right)\right]
 \hat{S}^{1/2}\psi_{n}(0),
 \label{TD-USPP-Immobile-Propagator}
\end{equation}
with $\hat{\cal T}$ the time-ordering operator. Algebraic expansions
of different order are then performed on the above propagator, leading
to various numerical time-stepping algorithms.

\subsection{First-order Implicit Euler Integration Scheme}

To first-order accuracy in time there are two well-known propagation
algorithms, namely, the explicit (forward) Euler
\begin{equation}
i\hbar\hat{S} \frac{ \psi_{n}(t+\Delta
  t) - \psi_{n}(t) }{\Delta t}
  = \hat{H} \psi_{n}({\bf
  x},t)
 \label{ExplicitEuler}
\end{equation}
and implicit (backward) Euler
\begin{equation}
i\hbar\hat{S} \frac{ \psi_{n}(t+\Delta t) - \psi_{n}({\bf
  x},t) }{\Delta t} = \hat{H} \psi_{n}(t+ \Delta t)
 \label{ImplicitEuler}
\end{equation}
schemes. Although the explicit scheme (\ref{ExplicitEuler}) is less
expensive computationally, our test runs indicate that it always
diverges numerically. The reason is that (\ref{TD-USPP-Immobile}) has
poles on the imaginary axis, which are marginally outside of the
stability domain (${\rm Re}(z\Delta t)<0$) of the explicit
algorithm. Therefore only the implicit algorithm can be used, which
upon rearrangement is,
\begin{equation}
\left[\hat{S} + \frac{i}{\hbar} \hat{H}\Delta t\right]
\psi_{n}(t+\Delta t) = \hat{S} \psi_{n}(t).
 \label{ImplicitEuler-Rearranged}
\end{equation}
In the above, we still have the choice of whether to use $\hat{H}(t)$
or $\hat{H}(t+\Delta t)$. Since this is a first-order algorithm,
neither choice would influence the order of the local truncation
error. Through numerical tests we found that the implicit time
differentiation in (\ref{ImplicitEuler}) already imparts
sufficient stability that the $\hat{H}(t+\Delta t)$ operator is not
needed. Therefore we will solve
\begin{equation}
  \left[\hat{S} + \frac{i}{\hbar} \hat{H}(t) \Delta t\right]
  \psi_{n}(t+\Delta t) = \hat{S} \psi_{n}(t)
 \label{ImplicitEuler-Rearranged-Final}
\end{equation}
at each timestep. Direct inversion turns out to be computationally
infeasible in large-scale planewave calculations. We solve
(\ref{ImplicitEuler-Rearranged-Final}) iteratively using matrix-free
linear equation solvers such as the conjugate gradient method.
Starting from the wavefunction of a previous timestep, we find that
typically it takes about three to five conjugate gradient steps to
achieve sufficiently convergent update.

One serious drawback of this algorithm is that norm conservation of
the wavefunction
\begin{equation}
 \langle \psi_{n}(t+\Delta t) | \hat{S} | \psi_{n}(t+\Delta t)
 \rangle \;=\;
 \langle \psi_{n}(t) | \hat{S} | \psi_{n}(t) \rangle
 \label{TD-USPP-Norm-Conservation}
\end{equation}
is not satisfied exactly, even if there is perfect floating-point
operation accuracy. So one has to renormalize the wavefunction after
several timesteps.

\subsection{First-order Crank-Nicolson Integration Scheme}

We find the following Crank-Nicolson expansion
\cite{CrankN47,KooninM89,CastroMR04} of propagator
(\ref{TD-USPP-Immobile-Propagator})
\begin{equation}
{\hat{S}}^{\frac{1}{2}} \psi_{n}(t+\Delta t) =
 \frac{1-\frac{i}{2\hbar}{\hat{S}}^{-\frac{1}{2}}\hat{H}(t)
 {\hat{S}}^{-\frac{1}{2}}\Delta t}{1+ \frac{i}{2\hbar}
   {\hat{S}}^{-\frac{1}{2}}\hat{H}(t) {\hat{S}}^{-\frac{1}{2}} \Delta t}
 {\hat{S}}^{\frac{1}{2}} \psi_{n}(t)
 \label{TD-USPP-First-Order-Crank-Nicolson-Expansion}
\end{equation}
stable enough for practical use. The norm of the wavefunction is
conserved explicitly in the absence of roundoff errors, because of the spectral
identity
\begin{equation}
\left\Vert\frac{1-\frac{i}{2\hbar} {\hat{S}}^{-\frac{1}{2}}\hat{H}
 {\hat{S}}^{- \frac{1}{2} }\Delta t}{1+ \frac{i}{2\hbar}
 {\hat{S}}^{- \frac{1}{2} }\hat{H} {\hat{S}}^{-\frac{1}{2} }\Delta t}
 \right\Vert = 1.
\end{equation}
Therefore (\ref{TD-USPP-Norm-Conservation}) is satisfied in an ideal
numerical computation, and in practice one does not have to
renormalize the wavefunctions in thousands of timesteps.

Writing out the (\ref{TD-USPP-First-Order-Crank-Nicolson-Expansion})
expansion explicitly, we have:
\begin{equation}
\left[\hat{S}+ \frac{i}{2\hbar} \hat{H}(t) \Delta t \right]\psi_{n}
    (t+\Delta t) = \left[ \hat{S} - \frac{i}{2\hbar} \hat{H}(t) \Delta
    t\right] \psi_{n}(t).
 \label{TD-USPP-First-Order-Crank-Nicolson}
\end{equation}
Similar to (\ref{ImplicitEuler-Rearranged-Final}), we solve
Eq. (\ref{TD-USPP-First-Order-Crank-Nicolson}) using the conjugate
gradient linear equations solver. This algorithm is still first-order
because we use $\hat{H}(t)$, not $(\hat{H}(t)+\hat{H}(t+\Delta t))/2$,
in (\ref{TD-USPP-First-Order-Crank-Nicolson}). In the limiting case of
time-invariant charge density, $\rho({\bf x},t)=\rho({\bf x},0)$ and
$\hat{H}(t+\Delta t)=\hat{H}(t)$, the algorithm has second-order
accuracy. This may happen if there is no external perturbation and we
are simply testing whether the algorithm is stable in maintaining the
eigenstate phase oscillation: $\psi_{n}(t)=\psi_{n}(0)e^{-i\omega t}$,
or in the case of propagating a test electron, which carries an
infinitesimal charge and would not perturb $\hat{H}(t)$.

\subsection{Second-order Crank-Nicolson Integration Scheme}

We note that replacing $\hat{H}(t)$ by $(\hat{H}(t)+\hat{H}(t+\Delta
t))/2$ in (\ref{TD-USPP-First-Order-Crank-Nicolson-Expansion}) would
enhance the local truncation error to second order, while still
maintaining norm conservation. In practice we of course do not know
$\hat{H}(t+\Delta t)$ exactly, which depends on $\rho(t+\Delta t)$ and
therefore $\psi_{n}(t+\Delta t)$.  However a sufficiently accurate
estimate of $\rho(t+\Delta t)$ can be obtained by running
(\ref{TD-USPP-First-Order-Crank-Nicolson}) first for one step, from
which we can get:
\begin{equation}
 \rho^\prime(t+\Delta t) \;=\; \rho(t+\Delta t) + {\cal O}(\Delta t^2),
\;\;
 \hat{H}^\prime(t+\Delta t) \;=\; \hat{H}(t+\Delta t) + {\cal
 O}(\Delta t^2).
\end{equation}
After this ``predictor'' step, we can solve:
\begin{equation}
 \left[\hat{S}+ \frac{i(\hat{H}(t)+\hat{H}^\prime(t+\Delta t)) \Delta
   t}{4\hbar} \right]\psi_{n} (t+\Delta t) = \left[ \hat{S} -
   \frac{i(\hat{H}(t)+\hat{H}^\prime(t+\Delta t)) \Delta t}{4\hbar}
   \right] \psi_{n}(t),
 \label{TD-USPP-Second-Order-Crank-Nicolson}
\end{equation}
to get the more accurate, second-order estimate for $\psi_{n}(t+\Delta
t)$, that also satisfies (\ref{TD-USPP-Norm-Conservation}).

\section{Optical Absorption Spectra}

Calculating the optical absorption spectra of molecules, clusters and
solids is one of the most important applications of TDDFT
\cite{ZangwillS80,BauernschmittA96,CasidaJCS98,YabanaB96,YabanaB99,
BertschIRY00,MarquesCR01,MarquesCBR03,TsolakidisSM02,OnidaRR02}. Since
many experimental and standard TDLDA results are available for
comparison, we compute the spectra for sodium dimer (${\rm Na}_2$) and
benzene molecule (${\rm C}_6{\rm H}_6$) to validate our direct
time-stepping USPP-TDDFT scheme.

We adopt the method by Bertsch {\em et al.}
\cite{YabanaB96,MarquesCR01} whereby an impulse electric field ${\bf
E}(t)=\epsilon\hbar\hat{\bf k}\delta(t)/e$ is applied to the system at
$t=0$, where $\hat{\bf k}$ is unit vector and $\epsilon$ is a small
quantity. The system, which is at its ground state at $t=0^-$, would
undergo transformation
\begin{equation}
 \tilde{\psi}_n({\bf x},t=0^+) \;=\; e^{i\epsilon\hat{\bf k}\cdot {\bf x}} 
   \tilde{\psi}_n({\bf x},t=0^-),
 \label{Impulse}
\end{equation}
for all its occupied electronic states, $n=1..N$, at $t=0^+$. Note
that the true, unpseudized wavefunctions should be used in
(\ref{Impulse}) if theoretical rigor is to be maintained.

One may then evolve $\{\tilde{\psi}_n({\bf x},t),n=1..N\}$ using a
time stepper, with the total charge density $\rho({\bf x},t)$ updated
at every step. The electric dipole moment ${\bf d}(t)$ is calculated
as
\begin{equation}
 {\bf d}(t) \;=\; e \int d^3{\bf x} \rho({\bf x},t) {\bf x}.
\end{equation}
In a supercell calculation one needs to be careful to have a large
enough vacuum region surrounding the molecule at the center, so no
significant charge density can ``spill over'' the PBC boundary, thus
causing a spurious discontinuity in ${\bf d}(t)$.

The dipole strength tensor ${\bf S}(\omega)$ can be computed by
\begin{equation}
 {\bf S}(\omega) \hat{\bf k} \;=\; {\bf m}(\omega) \equiv
 \frac{2m_e\omega}{e\hbar\pi} \lim_{\epsilon,\gamma\rightarrow 0}
 \frac{1}{\epsilon} \int_0^{\infty} dt \sin(\omega t) e^{-\gamma
 t^2}[{\bf d}(t) - {\bf d}(0)],
 \label{Response}
\end{equation}
where $\gamma$ is a small damping factor and $m_e$ is the electron
mass. In reality, the time integration is truncated at $t_{\rm f}$,
and $\gamma$ should be chosen such that $e^{-\gamma t_{\rm f}^2}\ll
1$. The merit of this and similar time-stepping approaches
\cite{LiY97} is that the entire spectrum can be obtained from just one
calculation.

For a molecule with no symmetry, one needs to carry out Eq.
(\ref{Impulse}) with subsequent time integration for three independent
$\hat{\bf k}$'s: $\hat{\bf k}_1, \hat{\bf k}_2, \hat{\bf k}_3$, and
obtain three different ${\bf m}_1(\omega), {\bf m}_2(\omega), {\bf
m}_3(\omega)$ on the right-hand side of Eq. (\ref{Response}). One then
solves the matrix equation:
\begin{equation}
 {\bf S}(\omega) [\hat{\bf k}_1 \; \hat{\bf
k}_2 \; \hat{\bf k}_3] \;=\; [{\bf m}_1(\omega) \; {\bf
m}_2(\omega) \; {\bf m}_3(\omega)] \;\;\rightarrow\;\; 
{\bf S}(\omega) \;=\; [{\bf m}_1(\omega) \; {\bf
m}_2(\omega) \; {\bf m}_3(\omega)] [\hat{\bf k}_1 \; \hat{\bf
k}_2 \; \hat{\bf k}_3]^{-1}.
\end{equation}
${\bf S}(\omega)$ satisfies the Thomas-Reiche-Kuhn $f$-sum rule,
\begin{equation}
 N\delta_{ij} \;=\; \int_0^{\infty} d\omega S_{ij}(\omega).
 \label{ThomasReicheKuhn}
\end{equation}
For gas-phase systems where the orientation of the molecule or cluster
is random, the isotropic average of ${\bf S}(\omega)$
\begin{equation}
 S(\omega) \;\equiv\; \frac{1}{3} {\rm Tr} {\bf S}(\omega)
\end{equation}
may be calculated and plotted.

In actual calculations employing norm-conserving pseudopotentials
\cite{MarquesCBR03}, the pseudo-wavefunctions ${\psi}_n({\bf x},t)$
are used in (\ref{Impulse}) instead of the true wavefunctions. And so
the oscillator strength ${\bf S}(\omega)$ obtained is not formally
exact. However, the $f$-sum rule Eq. (\ref{ThomasReicheKuhn}) is still
satisfied exactly. With the USPP/PAW formalism
\cite{Vanderbilt90,LaasonenCLV91,LaasonenPCLV93,Blochl94}, formally we
should solve
\begin{equation}
 \hat{T} {\psi_n}({\bf x},t=0^+) \;=\; e^{i\epsilon\hat{\bf k}\cdot {\bf x}} 
   \hat{T} {\psi_n}({\bf x},t=0^-), 
 \label{USPP-Perturbation}
\end{equation}
using linear equation solver to get ${\psi_n}({\bf x},t=0^+)$, and
then propagate ${\psi_n}({\bf x},t)$. However, for the present paper
we skip this step, and replace $\tilde{\psi}_n$ by ${\psi_n}$ in
(\ref{Impulse}) directly. This ``quick-and-dirty fix'' makes the
oscillator strength not exact and also breaks the sum rule
slightly. However, the peak positions are still correct.


\begin{figure}[th]
\includegraphics[width=5in]{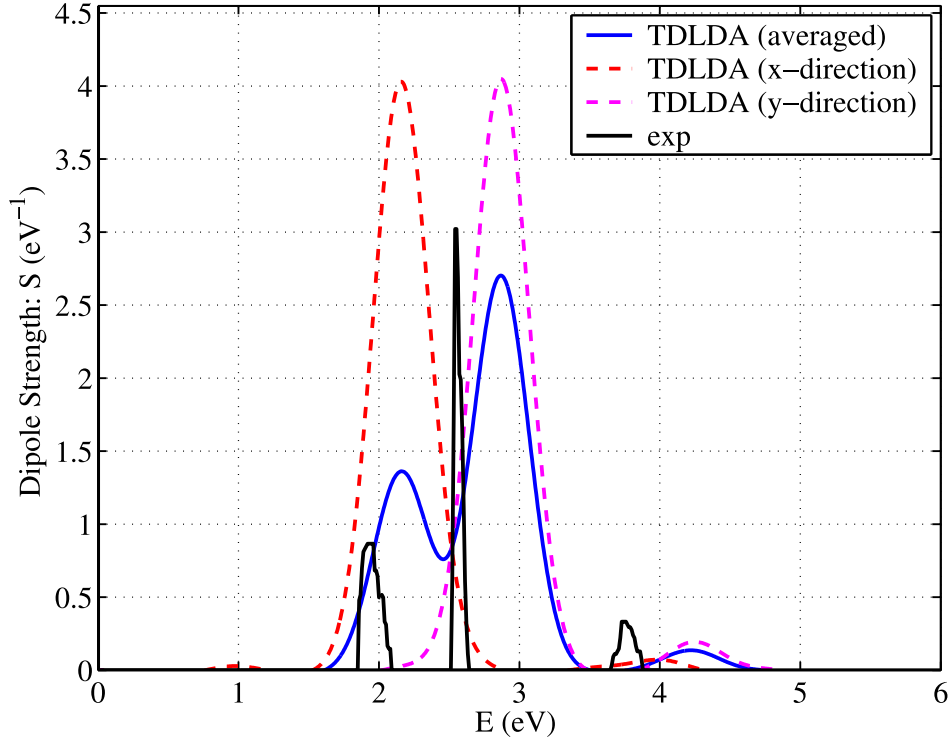}
\caption{Optical absorption spectra of ${\rm Na}_2$ cluster obtained
from direct time-stepping TDLDA calculation using norm-conserving TM
pseudopotential. The results should be compared with Fig. 1 of Marques
et al.  \cite{MarquesCR01}.}
\label{Sodium2_Spectrum}
\end{figure}

For the ${\rm Na}_2$ cluster, we actually use norm-conserving TM
pseudopotential \cite{DACAPO} for the Na atom, which is a special
limiting case of our USPP-TDDFT code. The supercell is a tetragonal
box of $12\times10\times10 \;{\rm \AA}^3$ and the ${\rm Na}_2$ cluster
is along the $x$-direction with a bond length of $3.0 \;{\rm
\AA}$. The planewave basis has a kinetic energy cutoff of $300$
eV. The time integration is carried out for $10,000$ steps with a
timestep of $\Delta t=1.97$ attoseconds, and $\epsilon=0.01/{\rm \AA}$,
$\gamma=0.02{\rm eV}^2/\hbar^2$. In the dipole strength plot
(Fig. \ref{Sodium2_Spectrum}), the three peaks agree very well with
TDLDA result from Octopus \cite{MarquesCR01}, and differ by $\sim 0.4$
eV from the experimental peaks \cite{Sinha49, FredricksonW27}.  In
this case, the $f$-sum rule is verified to be satisfied to within
$0.1\%$ numerically.

\begin{figure}[th]
\includegraphics[width=5in]{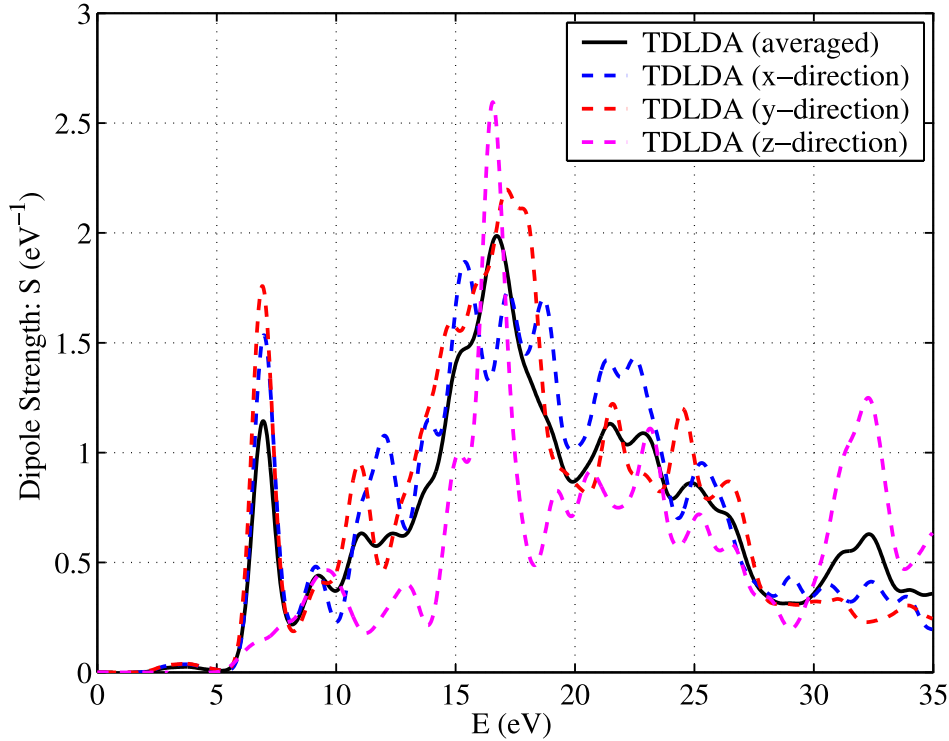}
\caption{Optical absorption spectrum of benzene (${\rm C}_6{\rm H}_6$)
molecule. The results should be compared with Fig. 2 of Marques et al.
\cite{MarquesCBR03}}
\label{Benzene_Spectrum}
\end{figure}

For the benzene molecule, ultrasoft pseudopotentials are used for both
carbon and hydrogen atoms. The calculation is performed in a
tetragonal box of $12.94\times10\times7 \;{\rm \AA}^3$ with the
benzene molecule placed on the $x-y$ plane. The C-C bond length is
$1.39 \;{\rm \AA}$ and the C-H bond length is $1.1 \;{\rm \AA}$. The
kinetic energy cutoff is $250$ eV, $\epsilon=0.01/{\rm \AA}$,
$\gamma=0.1{\rm eV}^2/\hbar^2$, and the time integration is carried
out for $5000$ steps with a timestep of $\Delta t=2.37$
attoseconds. In the dipole strength function plot
(Fig. \ref{Benzene_Spectrum}), the peak at $6.95$ eV represents the
$\pi\rightarrow\pi^*$ transition and the broad peak above $9$ eV
corresponds to the $\sigma\rightarrow\sigma^*$ transition. The dipole
strength function agrees very well with other TDLDA calculations
\cite{YabanaB99, MarquesCBR03} and experiment \cite{KochO72}. The
slight difference is mostly due to our {\em ad hoc} approximation that
${\psi}_n$'s instead of $\tilde{\psi}_n$'s are used in
(\ref{Impulse}). The more formally rigorous implementation of the
electric impulse perturbation, Eq. (\ref{USPP-Perturbation}), will be
performed in future work.

In this section we have verified the soundness of our time stepper
with planewave basis through two examples of explicit electronic
dynamics, where the charge density and effective potential are updated
at every timestep, employing both norm-conserving and ultrasoft
pseudopotentials. This validation is important for the following
non-perturbative propagation of electrons in more complex systems.

\section{Fermi Electron Transmission}

\begin{figure}[th]
 \centerline{\includegraphics[width=5in]{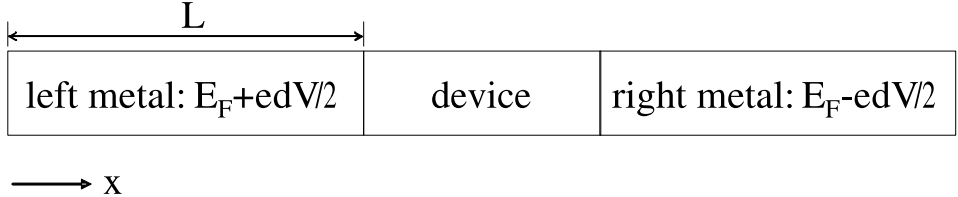}}
 \caption{Illustration of the Landauer transmission formalism.}
 \label{LandauerIllustration}
\end{figure}

We first briefly review the setup of the Landauer transmission
equation, \cite{Landauer57,Landauer70,ImryL99} before performing an
explicit TDDFT simulation. In its simplest form, two identical
metallic leads (see Fig. (\ref{LandauerIllustration})) are connected
to a device. The metallic lead is so narrow in $y$ and $z$ that only
one channel (lowest quantum number in the $y,z$ quantum well) needs to
be considered. In the language of band structure, this means that one
and only one branch of the 1D band structure crosses the Fermi level
$E_{\rm F}$ for $k_x>0$. Analogous to the universal density of states
expression $dN=2\Omega dk_xdk_ydk_z/(2\pi)^3$ for 3D bulk metals,
where $\Omega$ is the volume and the factor of $2$ accounts for up-
and down-spins, the density of state of such 1D system is simply
\begin{equation}
 dN \;=\; \frac{2L dk_x}{2\pi}.
\end{equation}
In other words, the number of electrons per unit length with
wave vector $\in(k_x, k_x+dk_x)$ is just $dk_x/\pi$. These electrons
move with group velocity \cite{Peierls55}:
\begin{equation}
 v_{\rm G} \;=\; \frac{dE(k_x)}{\hbar dk_x},
 \label{GroupVelocity}
\end{equation}
so there are $(dk_x/\pi)(dE(k_x)/(\hbar dk_x))=2dE/h$
such electrons hitting the device from either side per unit time.

Under a small bias voltage $dV$, the Fermi level of the left lead is
raised to $E_{\rm F}+edV/2$, while that of the right lead drops to
$E_{\rm F}-edV/2$. The number of electrons hitting the device from the
left with wave vector $(k_x, k_x+dk_x)$ is exactly equal to the number
of electrons hitting the device from the right with wave vector
$(-k_x, -k_x-dk_x)$, except in the small energy window $(E_{\rm
F}-edV/2,E_{\rm F}+edV/2)$, where the right has no electrons to
balance against the left. Thus, a net number of $2(edV)/h$ electrons
will attempt to cross from left and right, whose energies are very
close to the original $E_{\rm F}$. Some of them are scattered back by
the device, and only a fraction of $T\in(0,1]$ gets through. So the
current they carry is:
\begin{equation}
 \left.\frac{dI}{dV}\right|_{V=0} \;=\; \frac{2e^2}{h}T(E_{\rm F}),
 \label{LandauerFormula}
\end{equation}
where ${2e^2}/{h}=77.481\mu S=(12.906 k\Omega)^{-1}$.

Clearly, if the device is also of the same material and structure as
the metallic leads, then $T(E_{\rm F})$ should be $1$, 
when we ignore electron-electron and electron-phonon scattering. This
can be used as a sanity check of the code. For a nontrivial device
however such as a molecular junction, $T(E_{\rm F})$ would be smaller
than $1$, and would sensitively depend on the alignment of the
molecular levels and $E_{\rm F}$, as well as the overlap between these
localized molecular states and the metallic states.

\begin{figure}[th]
 \subfigure[]{\includegraphics[width=5in]{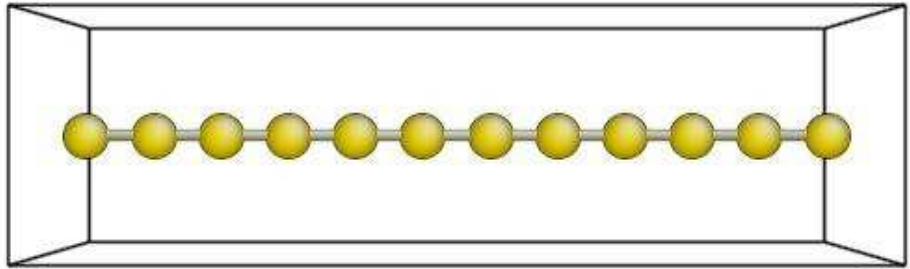}}
 \subfigure[]{\includegraphics[width=5in]{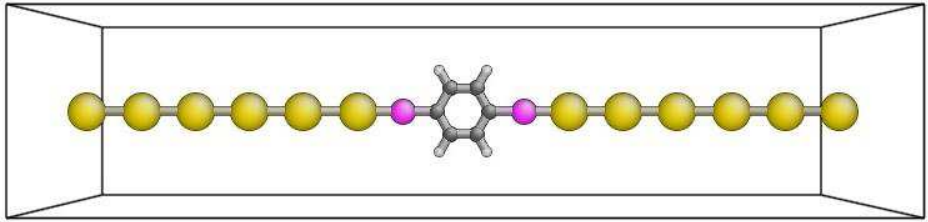}}
 \caption{Atomistic configurations of our USPP-TDDFT simulations (Au:
 yellow, S: magenta, C: black, and H: white). (a) 12-atom Au
 chain. Bond length: Au-Au 2.88 $\;{\rm \AA}$. (b) BDT (-S-C$_6$H$_4$-S-)
 junction connected to Au chain contacts. Bond lengths: Au-Au 2.88
 {\AA}, Au-S 2.41 {\AA}, S-C 1.83 {\AA}, C-C 1.39 {\AA}, and C-H 1.1
 {\AA}.}
\label{Configuration1}
\end{figure}

Here we report two USPP-TDDFT case studies along the line of the above
discussion. One is an infinite defect-free gold chain
(Fig. \ref{Configuration1}(a)). The other case uses gold chains as
metallic leads and connects them to a -S-C$_6$H$_4$-S-
(benzene-(1,4)-dithiolate, or BDT) molecular junction
(Fig. \ref{Configuration1}(b)).

In the semi-classical Landauer picture explained above, the metallic
electrons are represented by very wide Gaussian wavepacks
\cite{Peierls55} moving along with the group velocity $v_{\rm G}$, and
with negligible rate of broadening compare to $v_{\rm G}$. Due to
limitation of computational cost, we can only simulate rather small
systems. In our experience with 1D lithium and gold chains, a Gaussian
envelop of 3-4 lattice constants in full width half maximum is
sufficient to propagate at the Fermi velocity $v_{\rm G}(k_{\rm F})$
with 100\% transmissions and maintain its Gaussian-profile envelop
with little broadening for several femto-seconds.

\subsection{Fermi electron propagation in gold chain}

The ground-state electronic configurations of pure gold chains are
calculated using the free USPP-DFT package DACAPO,
\cite{DACAPO,HammerHN99,BahnJ02} with local density functional (LDA)
\cite{CeperleyA80,PerdewZ81} and planewave kinetic energy cutoff of
$250$ eV. The ultrasoft pseudopotential is generated using the free
package uspp (ver. 7.3.3)
\cite{Vanderbilt90,LaasonenCLV91,LaasonenPCLV93}, with $5d$, $6s$,
$6p$, and auxiliary channels. Fig. \ref{Configuration1}(a) shows a
chain of 12 Au atoms in a tetragonal supercell ($34.56\times 12\times
12$ {\AA}$^3$), with equal Au-Au bond length of $2.88$
{\AA}. Theoretically, 1D metal is always unstable against
period-doubling Peierls distortion \cite{Peierls55,Marder00}. However,
the magnitude of the Peierls distortion is so small in the Au chain
that room-temperature thermal fluctuations will readily erase its
effect. For simplicity, we constrain the metallic chain to maintain
single periodicity. Only the $\Gamma$-point wavefunctions are
considered for the 12-atom configuration.

\begin{figure}[th]
\includegraphics[width=5in]{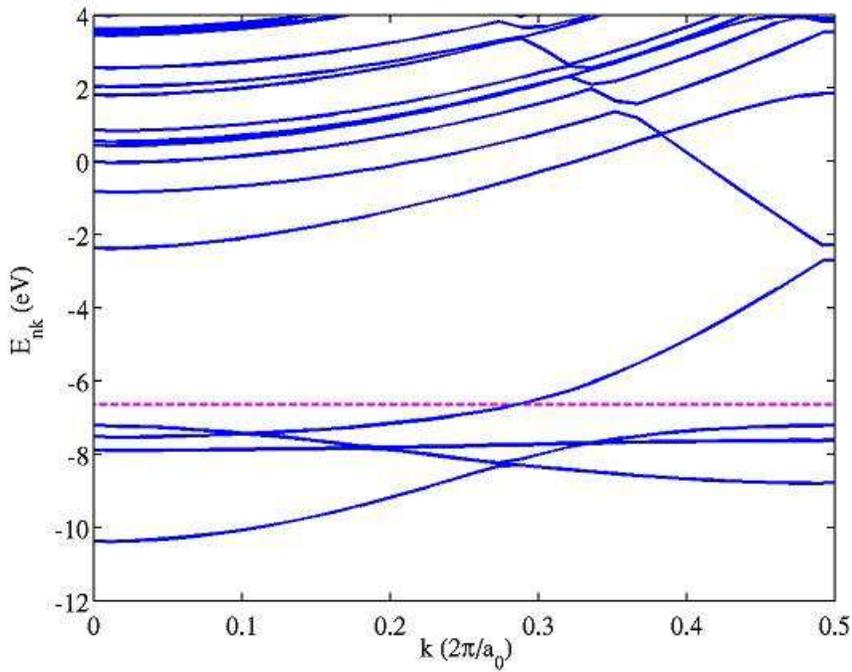}
\caption{Band structure of a one-atom Au chain with 64
 Monkhorst-Pack\cite{MonkhorstP76} {\bf k}-sampling in the chain
 direction. The Fermi level, located at $-6.65$ eV, is marked as the
 dashed line.}
\label{GoldChain_BandStructure}
\end{figure}

The Fermi level $E_{\rm F}$ is found to be $-6.65$ eV, which is
confirmed by a more accurate calculation of a one-Au-atom system with
{\bf k}-sampling (Fig. \ref{GoldChain_BandStructure}). The Fermi state
is doubly degenerate due to the time-inversion symmetry, corresponding
to two Bloch wavefunctions of opposite wave vectors $k_{\rm F}$ and
$-k_{\rm F}$.

From the $\Gamma$-point calculation, two energetically degenerate and
real eigen-wavefunctions, $\psi_+({\bf x})$ and $\psi_{-}({\bf x})$,
are obtained. The complex traveling wavefunction is reconstructed as
\begin{equation}
 \psi_{k_{\rm F}}({\bf x}) = \frac{\psi_+({\bf x}) + i \psi_{-}({\bf
x})}{\sqrt{2}}.
\end{equation}
The phase velocity of ${\psi}_{k_{\rm F}}({\bf x},t)$ computed from
our TDLDA runs matches the Fermi frequency $E_{\rm F}/\hbar$. We use
the integration scheme (\ref{TD-USPP-First-Order-Crank-Nicolson}) and
a timestep of $2.37$ attoseconds.

\begin{figure}[th]
\includegraphics[width=5in]{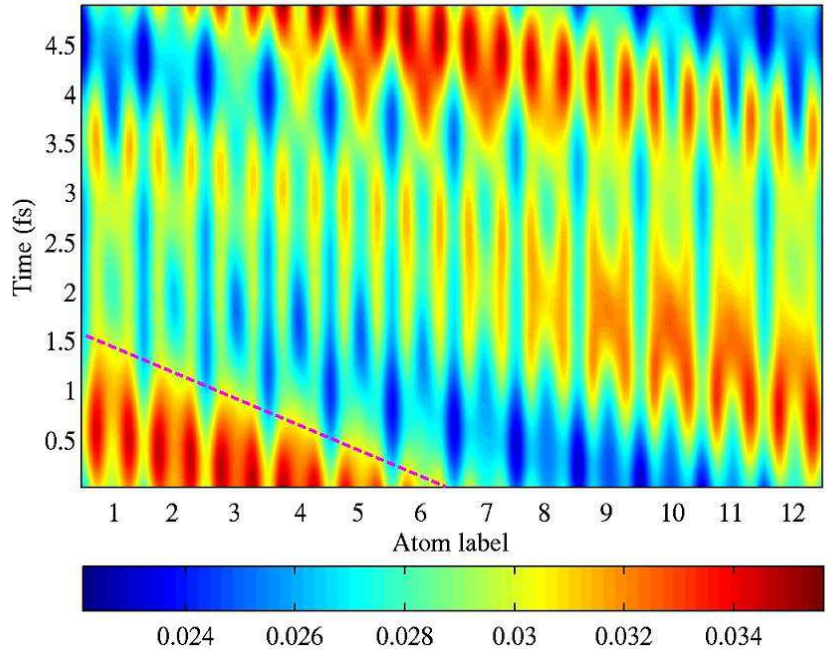}
\caption{Evolution of modulated Fermi electron density in time along
  the chain direction. The electron density, in the unit of ${\rm
  \AA}^{-1}$, is an integral over the perpendicular $y$-$z$ plane and
  normalized along the $x$ direction, which is then color coded.}
\label{GoldChain_Propagation}
\end{figure}

We then calculate the Fermi electron group velocity $v_{\rm G}(k_{\rm
F})$ by adding a perturbation modulation of
\begin{equation}
 \widetilde{\psi}_{k_{\rm F}}({\bf x},t=0) \;=\; \psi_{k_{\rm F}}({\bf x})
 (1 + \lambda\sin(2\pi x/L))
\end{equation}
to the Fermi wavefunction ${\psi}_{k_{\rm F}}({\bf x})$, where
$\lambda$ is $0.02$ and $L$ is the $x$-length of the
supercell. Fig. \ref{GoldChain_Propagation} shows the electron density
plot along two axes, $x$ and $t$. From the line connecting the
red-lobe edges, one can estimate the Fermi electron group velocity to
be $\sim$10.0 \AA/fs. The Fermi group velocity can also be obtained
analytically from Eq. (\ref{GroupVelocity}) at $k_x=k_{\rm F}$. A
value of 10 \AA/fs is found according to
Fig. \ref{GoldChain_BandStructure}, consistent with the TDLDA result.

\begin{figure}[th]
\includegraphics[width=5in]{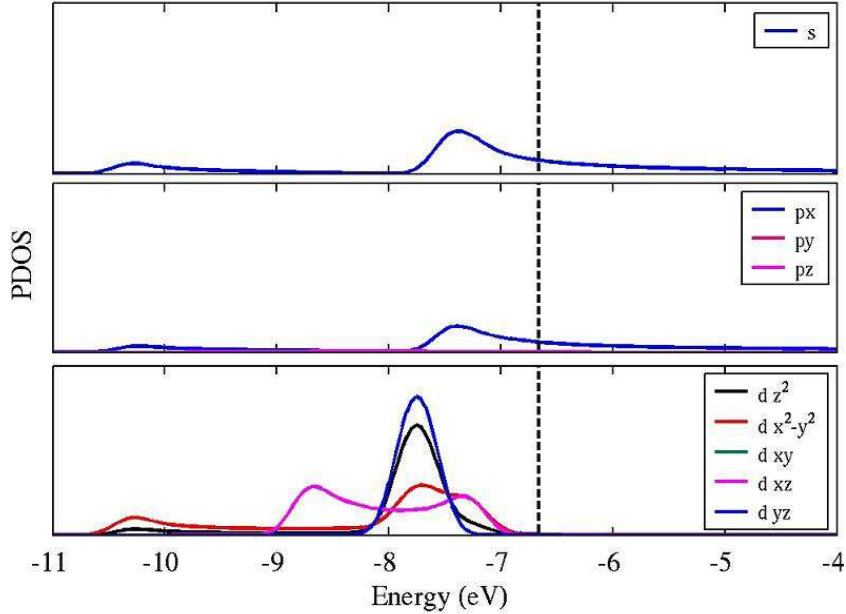}
\caption{Projected density of states of the 12-atom Au chain.}
\label{GoldChain_PDOS}
\end{figure}

Lastly, the angular momentum projected densities of states are shown
in Fig. \ref{GoldChain_PDOS}, which indicate that the Fermi
wavefunction mainly has $s$ and $p_x$ characteristics.

\subsection{Fermi electron transmission through Au-BDT-Au junction}

At small bias voltages, the electric conductance of a molecular
junction (Fig. \ref{Configuration1}(b)) is controlled by the
transmission of Fermi electrons, as shown in
Eq. (\ref{LandauerFormula}). In this section, we start from the Fermi
electron wavefunction of a perfect 1D gold chain
(Fig. \ref{Configuration1}(a)), and apply a Gaussian window centered
at ${\bf x}_0$ with a half width of $\sigma$, to obtain a localized
wave pack
\begin{equation}
 \widetilde{\psi}_{k_{\rm F}}({\bf x},t=0) = {\psi}_{k_{\rm F}}({\bf
 x}) G\left(\frac{{\bf x}-{\bf x}_0}{\sigma}\right),
\end{equation}
at the left lead. This localized Fermi electron wave pack is then
propagated in real time by the TDLDA-USPP algorithm
(\ref{TD-USPP-First-Order-Crank-Nicolson}) with a timestep of $2.37$
attoseconds, leaving from the left Au lead and traversing across the
-S-C$_6$H$_4$-S- molecular junction (Fig. \ref{Configuration1}(b)).
While crossing the junction the electron will be scattered, after
which we collect the electron density entering the right Au lead to
compute the transmission probability $T(E_{\rm F})$ literally. The
calculation is performed in a tetragonal box ($42.94\times 12\times
12$ {\AA}$^3$) with a kinetic energy cutoff of $250$ eV.

\begin{figure}[th]
\includegraphics[width=5in]{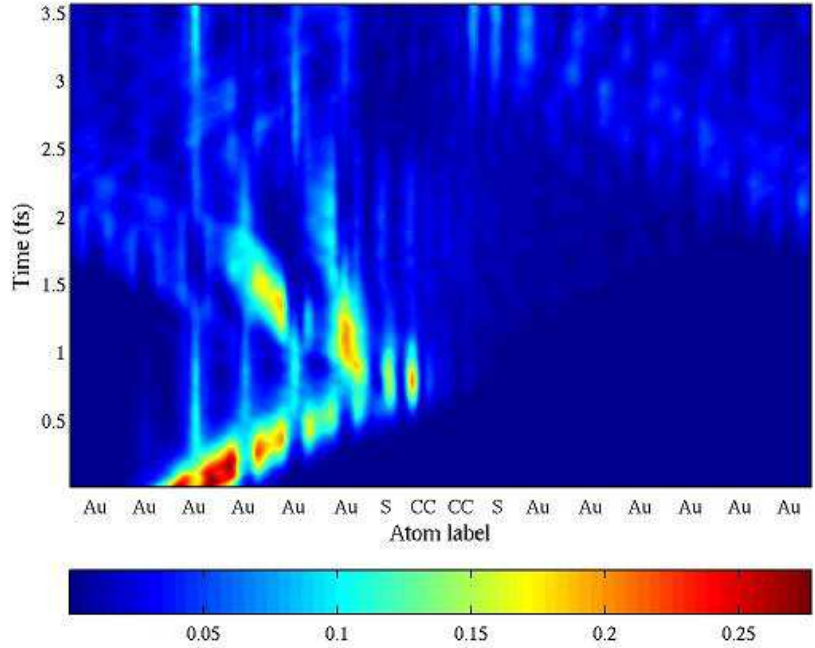}
\caption{Evolution of filtered wave package density in time along the
  chain direction. The electron density, in the unit of ${\rm
  \AA}^{-1}$, is a sum over the perpendicular $y$-$z$ plane and normalized
  along the $x$ direction. The normalized electron density is color
  coded by the absolute value.}
\label{GoldJunction_Propagation}
\end{figure}

Fig. \ref{GoldJunction_Propagation} shows the Fermi electron density
evolution in $x$-$t$.  A group velocity of $10$ {\AA}/fs is obtained
from the initial wave pack center trajectory, consistent with the
perfect Au chain result. This {\it free} propagation lasts for about
$0.8$ fs, followed by a sharp density turnover that indicates the
occurrence of strong electron scattering at the junction.  A very
small portion of the wave pack goes through the molecule. After
about $1.7$ fs, the reflected portion of the wave pack enters the
right side of the supercell through PBC.

\begin{figure}[th]
\includegraphics[width=5in]{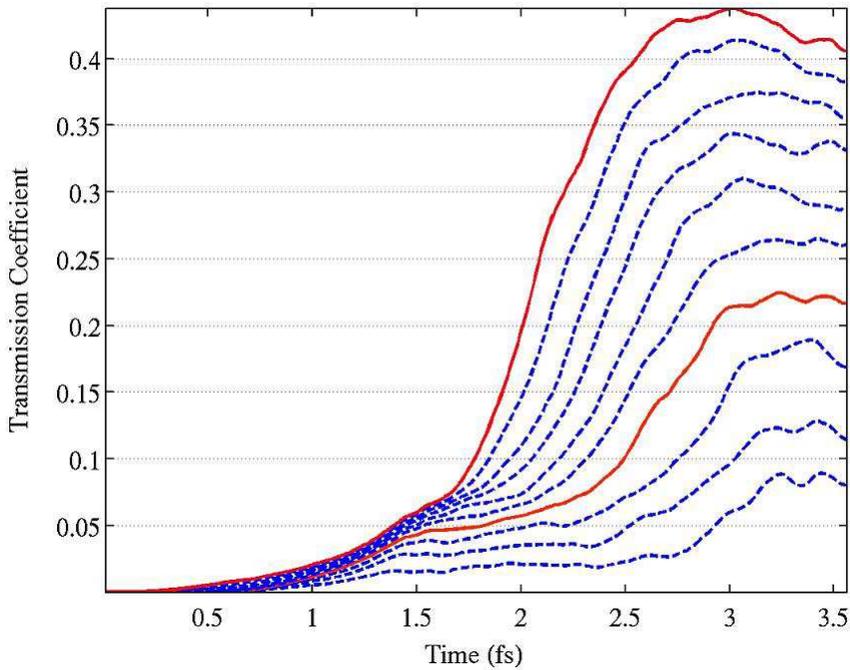}
\caption{$R(x^\prime,t)$ versus time plot. Curves are measured in $10$
  different regions with different $x^\prime$ positions, which equally
  divide the region from the right S atom to the boundary on the right
  hand side. } 
 \label{GoldJunction_Propagation_DirX_Cut}
\end{figure}

To separate the transmitted density from the reflected density as
clearly as possible, we define and calculate the following cumulative
charge on the right side
\begin{equation}
 R(x^\prime,t) \;\equiv\; \int_{x_{\rm S}}^{x^\prime}dx \int_0^{L_y} dy
 \int_0^{L_z} dz \rho(x,y,z,t),
\end{equation}
where $x_{\rm S}$ is the position of the right sulfur
atom. $R(x^\prime,t)$ is plotted in
Fig. \ref{GoldJunction_Propagation_DirX_Cut} for ten
$x^\prime$-positions starting from the right sulfur atom up to the
right boundary $L_x$. A shoulder can be seen in all 10 curves, at
$t=1.5$-$2$ fs, beyond which $R(x^\prime,t)$ starts to rise sharply
again, indicating that the reflected density has entered from the
right boundary. Two solid curves are highlighted in
Fig. \ref{GoldJunction_Propagation_DirX_Cut}. The lower curve is at
$x^\prime=x_{\rm S}+7.2$ {\AA}, which shows a clear transmission
plateau of about $5$\%. The upper curve, which is for $x^\prime$
exactly at the right PBC boundary, shows $R(x^\prime,t)\approx 7$\% at
the shoulder. From these two curves, we estimate a transmission
probability $T(E_{\rm F})$ of $5$-$7$\%, which corresponds to a
conductance of $4.0$-$5.6$ $\mu$S according to
Eq. (\ref{LandauerFormula}). This result from planewave TDLDA-USPP
calculation is comparable to the transmission probability estimate of
$10$\% from complex band structure calculation
\cite{TomfohrS02a,TomfohrS02b} for one benzene linker (-C$_6$H$_4$-)
without the sulfur atoms, and the non-equilibrium Green's function
estimate of $5$ $\mu$S \cite{XueR03} for the similar system.

\section{Summary}

In this work, we develop TDDFT based on Vanderbilt ultrasoft
pseudopotentials and benchmark this USPP-TDDFT scheme by calculating
optical absorption spectra, which agree with both experiments and
other TDDFT calculations. We also demonstrate a new approach to
compute the electron conductance through single-molecule junction via
wave pack propagation using TDDFT. The small conductance of
$4.0$-$5.6$ $\mu$S is a result of our fixed band approximation,
assuming the electron added was a small testing electron and therefore
generated little disturbing effects of the incoming electrons on the
electronic structure of the junction. This result is of the same order
of magnitude as the results given by the Green's function and the
complex band approaches, both requiring similar assumptions. 

\begin{acknowledgments}
We thank Peter Bl\"{o}chl for valuable suggestions. XFQ, JL and XL are
grateful for the support by ACS to attend the TDDFT 2004 Summer School
in Santa Fe, NM, organized by Carsten Ullrich, Kieron Burke and
Giovanni Vignale.  XFQ, XL and SY would like to acknowledge support by
DARPA/ONR, Honda R\&D Co., Ltd. AFOSR, NSF, and LLNL. JL would like to
acknowledge support by Honda Research Institute of America, NSF
DMR-0502711, AFOSR FA9550-05-1-0026, ONR N00014-05-1-0504, and the
Ohio Supercomputer Center.
\end{acknowledgments}


\end{document}